\pdfoutput=1
\documentclass[12pt]{article}
\usepackage[english]{babel}
\usepackage{color}
\usepackage{graphicx}
\usepackage{amsmath}
\usepackage{amsthm}
\usepackage{amssymb}
\usepackage[hyperindex,plainpages=false]{hyperref}
\usepackage[left=1.1in,right=1.1in,bottom=1in,top=1in,a4paper]{geometry}

\title{Symmetry structure of integrable hyperbolic third order equations}
\author{Alexander G. Rasin  \\
Department of Mathematics,\\ 
Ariel University, Ariel 40700, Israel \\
{E-mail: rasin@ariel.ac.il}\and  Jeremy Schiff  \\
Department of Mathematics,\\
Bar-Ilan University, Ramat Gan, 52900, Israel \\
{E-mail: schiff@math.biu.ac.il}}

\begin{document}
\maketitle
\begin{abstract}
  We explore the application of generating symmetries, i.e. symmetries that depend on a parameter, to
  integrable hyperbolic third order equations, and in particular to consistent pairs of such
  equations as introduced by Adler and Shabat in \cite{AS}. Our main result is that different infinite
  hierarchies of symmetries for these equations can arise from a single generating symmetry by
  expansion about different values of the parameter. We illustrate this, and study in depth
  the symmetry structure, for two examples. The first is an equation related to the potential KdV
  equation taken from \cite{AS}. The second is a more general hyperbolic equation than the kind considered
  in \cite{AS}. Both equations depend on a parameter, and when this parameter vanishes they become part
  of a consistent pair. When this happens, the nature of the expansions of the generating symmetries needed
  to derive the hierarchies also changes. 
\end{abstract}

\section{Introduction} 

Although the original characterization of integrability for partial differential equations (PDE)
was through  the existence of an infinite number of conservation laws  \cite{MGK}, it was soon realized that
integrable PDE also exhibit infinitely many infinitesimal symmetries \cite{Ol1}.  The latter property is
easier to investigate systematically, and has become a central pillar in the classification of 
integrable PDE. See \cite{book1,book2} for surveys, and the papers \cite{cp1,cp2,cp3}   for some examples
of applications to different classes of equations.

One of the important classes of equations to which the symmetry method has been applied \cite{cp4,cp5}
is the class of scalar, second order hyperbolic equations of the form  
\begin{equation}
u_{xy} = h(x,y,u,u_x,u_y) \ .
\label{2ord} \end{equation}   
(See also \cite{cp6,cp7} for vector extensions.)  
For equations of this type, the conventional wisdom is that integrable cases are characterized 
by {\em two} infinite hierarchies of symmetries.  In fact, a little more detail is needed in this discussion.
Even for the archetypical example of an integrable PDE, the Korteweg-de Vries (KdV) equation, there are two infinite
hierarchies of symmetries, the standard hierarchy of commuting symmetries, as discussed in \cite{Ol1}, and the
hierarchy of ``additional'' symmetries \cite{addsym1,addsym2,addsym3}, which do not commute, either among themselves or with the standard
symmetries. The latter are often ignored; this may be because with the exception of the two lowest order symmetries in
this hierarchy, they are nonlocal. But these two lowest order symmetries are the Galilean and scaling symmetries, which
certainly play  a significant role in the theory of the KdV equation. 
In greater generality, for many (and maybe all) integrable PDE it is possible to identify one or more
{\em recursion operators} that map symmetries to symmetries, and application of these recursion operators
to simple point symmetries can generate many hierarchies (see \cite{cp8} for a good example).
However, it seems that  integrable equations of the form (\ref{2ord}) are characterized by the existence of
at least two infinite hierarchies of local, {\em commuting} symmetries. 

In our previous works \cite{RSG,RS5,RS6,RS7,RS8,RSlast,Sashalast}  we showed that the infinite hierarchies of symmetries of
integrable PDE can be conveniently encapsulated in {\em generating symmetries}, which are non-local symmetries
depending on a parameter, that give rise to infinite hierarchies when expanded in suitable power
series in the parameter. In the current paper we extend this work to some third order hyperbolic PDEs. 
As opposed to second order hyperbolic PDEs, third order hyperbolic PDEs have hardly been studied.
An exception  is the ground-breaking paper of Adler and Shabat \cite{AS} on equations of the form 
\begin{equation}
u_{xxy} = f(x,y,u,u_x,u_y,u_{xy},u_{xx}) \ .
\label{3ord} \end{equation}   
In addition to giving various examples of integrable equations of this form, 
Adler and Shabat point out that there are examples of such equations which are consistent with another 
third order equation of the form 
\begin{equation}
u_{xyy} = g(x,y,u,u_x,u_y,u_{xy},u_{yy}) \ , 
\label{3o2} \end{equation}   
but are irreducible, in the sense that (\ref{3ord}) and (\ref{3o2})
are not obtained simply by differentiation, with respect to $x$ and $y$ respectively,  of a second order
equation of the form (\ref{2ord}). Adler and Shabat call such pairs of equations ``consistent pairs'' 
and point out that they ``belong to an intermediate class between those of the second and third order equations''.
An interesting application of Adler and Shabat's idea can be found in  \cite{DS}. From the point of
view of symmetries, although full details are not given in \cite{AS}, it seems that integrable equations of form 
(\ref{3ord}) also have two infinite hierarchies of local, commuting symmetries, but some kind of degeneration
takes place when the equation is part of a consistent pair.  

The contributions and structure of the current paper are as follows: in Section 2 we review the main definitions from \cite{AS},
and show that the notion of consistent pair can be extended to more general third order hyperbolic equations. In section 3, we reconsider the example
of Adler-Shabat that is related to the potential KdV equation, viz. the equation
\begin{equation}
  u_{xxy} =  \frac{ u_{xy}^2 - c }{2u_y}  + 2u_x u_y \label{ex}
\end{equation}   
where $c$ is a constant.  We give the Lax pair for this equation,  
in several forms,  write down generating symmetries, and show, remarkably, that {\em both} infinite hierarchies of local, commuting symmetries arise from
a {\em single} generating symmetry, via expansions around two different values of the parameter. In the case $c=0$,   
when the relevant equation is part of a consistent pair, we show that the dependence  of the generating symmetry on the
parameter changes, and thus different expansions are needed, but still the two hierarchies are obtained from the same generating symmetry,  via two
expansions around different values of the parameter. Thus the generating symmetry turns out to be a unifying object. In Section 4 we look
at a more intricate example,  viz. the equation
\begin{equation}
  u_{xxy}   =   \frac{u_{xy}^2 - c}{2u_y} -  u_yu_{xxx}     \label{ne1}
\end{equation}
that is not in the form (\ref{3ord}).  Once again, we give the Lax pair and write
down generating symmetries, but in this case it is necessary to consider expansions around three different values of the parameter,
and the resulting symmetries do not all commute. Once again, the form of the expansions change when $c=0$ and  the equation is part
of a consistent pair; in this case it is related by a differential substitution to the second order hyperbolic equation
\begin{equation}  w_{xy} + \left( w_{xx} + {\textstyle{\frac12}} w_x^2  \right) e^w + f e^{-w} = 0   \label{2hyp}   \end{equation}
where $f$ is a constant. But still the generating symmetry is a single unifying object that covers all the different cases and
encodes numerous different hierarchies. 
For equation (\ref{ne1}), we complete the theory by also looking at the generating symmetry that gives rise to additional symmetries
and its expansions, and we show the origin of various different recursion operators from certain identities
satisfied by the generating symmetries.  Section 5 contains some concluding remarks and open questions.  

We end this introduction by referring to the paper \cite{HK}, which introduces a "generalized invariant manifold" for some integrable equations, including KdV. The authors demonstrate (Section 4) that the generalized invariant manifold $W=\varphi^2$ is a symmetry of pKdV, where $\varphi$ is a solution of the Lax Pair. This symmetry was discovered in \cite{Dic,lou2012,OEVEL1998161}, and is characterized as a generating symmetry in \cite{RSG}. In another paper \cite{cp9}, the authors shows that for a second order hyperbolic equation, in the integrable case, two recursion operators (generating two distinct infinite hierarchies of symmetries) are produced by different parametrizations of the same generalized invariant manifold. This result coincides with the result that we obtain here.   

\section{Adler-Shabat Pairs and a Generalization}

In this section we briefly recap the main definitions given by Adler and Shabat in \cite{AS} and show how the notion of a consistent pair can be
generalized.  The following is taken almost verbatim from \cite{AS}:  
\begin{itemize}
\item Second order and third order hyperbolic equations are, respectively, equations of the form
(\ref{2ord}) and (\ref{3ord}).  
\item A pair of third order hyperbolic equations of the form (\ref{3ord}) and (\ref{3o2}). 
  is called a {\em consistent pair} if $D_y(f) = D_x(g)$ on solutions of the pair of equations, where $D_x,D_y$ are total derivative operators. 
\item A consistent pair of third order hyperbolic equations is called reducible if its general solution solves a
  1-parameter family of second order  hyperbolic equations $u_{xy} = h(\alpha;x,y,u,u_x,u_y)$. Otherwise it is called irreducible. 
\item  A second order or third order hyperbolic equation is integrable if it is compatible with an infinite
  hierarchy of evolutionary symmetries. 
\item A representation of the consistent pair (\ref{3ord})-(\ref{3o2})   in B\"acklund variables is a system 
  \begin{equation}\begin{array}{l} 
    v_{x} = F(x,y,u,u_x,v) \ , \\
    v_{y} = G(x,y,u,u_y,v) \ ,
    \end{array}     \label{Bsys}
    \end{equation}   
    from which the pair follows by elimination of $v$. (Cross differentiation and elimination of derivatives of $v$ yields
    a relation of the form $u_{xy} = H(x,y,u,u_x,u_y,v)$ and further differentiation with respect to each of the variables
    and elimination of $v$ yields the two equations of the pair.)   
\item A representation of the single third order hyperbolic equation (\ref{3ord}) in B\"acklund variables is a system 
  \begin{equation}\begin{array}{l} 
    v_{x} = F(x,y,u,u_x,v) \ , \\
    u_{xy} = H(x,y,u,u_x,u_y,v)       
    \end{array}     \label{Bsys2}
    \end{equation}   
from which the equation follows by elimination of $v$.  
\end{itemize} 

Quite clearly, equation (\ref{3ord}) is not the most general form for a third order hyperbolic equation.
For example, we might also  consider equations of either of the forms
\begin{equation}
  u_{xxy} +  a(u,u_x,u_y) u_{xxx}  =  f(u,u_x,u_y,u_{xx},u_{xy})\ ,   \label{3gen1}
\end{equation}
or
\begin{equation}
  u_{xyy} +  b(u,u_x,u_y) u_{xxx}  =  g(u,u_x,u_y,u_{xx},u_{xy},u_{yy})\ ,  \label{3gen2} 
\end{equation} 
where in the second case we require $b<0$ for hyperbolicity. We claim the following: {\em if $b=-a^2$ then for
suitable choices of $a,f,g$ the pair of equations (\ref{3gen1}) and (\ref{3gen2}) can be consistent.} 

To show this we proceed as follows.  Writing
$$ e_1 = u_{xxy} + au_{xxx} - f \ , \qquad  e_2 = u_{xyy} + bu_{xxx} - g\ ,  $$ 
consistency requires the existence of an identity of the form
\begin{equation}
D_y(e_1) - D_x(e_2) - a D_x(e_1) = \alpha_1 e_1 + \alpha_2 e_2 \label{id} 
\end{equation} 
where $\alpha_1,\alpha_2$ depend on $u$ and its first and second derivatives. 
The terms on the left hand side of  this identity were chosen so that the only fourth derivative of $u$ that appears is
$u_{xxxx}$. This appears linearly, with the coefficient $-b-a^2$, and thus it is necessary to choose $b=-a^2$. All the third derivatives of
$u$ also appear linearly (and $u_{yyy}$ does not appear at all). The functions $\alpha_1,\alpha_2$ can be chosen so that
the coefficients of $u_{xxy}$ and $u_{xyy}$ vanish, and for the coefficient of $u_{xxx}$ also to vanish we find that it
is necessary to impose the condition
\begin{equation}
  a \tilde{D}_x  \left(a \right)+\tilde{D}_y  \left(a \right)
  + a^{2} \left(  \frac{\partial g}{\partial u_{yy}} -2 \frac{\partial f}{\partial u_{xy}}  \right)
  + a \left( 2  \frac{\partial f}{\partial u_{xx}}  -\frac{\partial g}{\partial u_{xy}}   \right)
  +\frac{\partial g}{\partial u_{xx}}=0 \ ,   \label{cond1}  
\end{equation}
where 
\[
\tilde{D}_x(f)=u_x\frac{\partial f}{\partial u}+u_{xx}\frac{\partial f}{\partial u_x}+u_{xy}\frac{\partial f}{\partial u_y}\ ,~~~
\tilde{D}_y(f)=u_y\frac{\partial f}{\partial u}+u_{xy}\frac{\partial f}{\partial u_x}+u_{yy}\frac{\partial f}{\partial u_y}\ .
\]
Once this is done, the consistency condition reduces to a condition involving only $u$ and its first and second derivatives, viz. 
\begin{equation}
  \tilde{D}_y  \left(f \right)-\tilde{D}_x  \left(g \right) - a \tilde{D}_x  \left(f \right) +
  f\left(\frac{\partial f}{\partial u_{xx}}  -a \frac{\partial f}{\partial u_{xy}}-\frac{\partial g}{\partial u_{xy}}\right)
  +g\left(\frac{\partial f}{\partial u_{xy}}-\frac{\partial g}{\partial u_{yy}}\right)=0\ .
         \label{cond2}  
\end{equation}

We note the following:
\begin{itemize}
\item   In the case $a=\frac{\partial g}{\partial u_{xx}}=0$ 
  the condition (\ref{cond1}) is satisfied automatically, and the condition (\ref{cond2}) reduces to
  the requirement $D_y(f)=D_x(g)$ on solutions of the pair of equations (\ref{3ord})-(\ref{3o2}).
  Thus we recover the original consistent pairs from \cite{AS}. 
\item   Since, for given $e_1$, the consistency condition (\ref{id}) can be regarded as a first order differential equation for $e_2$,
  there is in fact always the possibility to insert an extra term with an arbitrary function of $y$ in the second equation of a pair of the form
  (\ref{3gen1})-(\ref{3gen2}).  We will see this in examples later on.  
\item   If  a pair (\ref{3gen1})-(\ref{3gen2}) can be obtained by differentation, with respect to $x$ and $y$, 
  of a single second order differential equation of
  the form  $ h(x,y,y,u_x,u_x,u_y,u_{xy},u_{xx})  = 0  $, we call it reducible. Otherwise, the pair is irreducible. 
\item We leave for further investigation the question of how many solutions exist of the constraints (\ref{cond1})-(\ref{cond2}).
  In this paper we focus on a single example (with $a=0$ and $g$ independent of $u_{xx}$) from \cite{AS}, 
  and a single  example of the new type with  
  $$  a= u_y \ , \qquad  f = \frac{u_{xy}^2}{2u_y}  \ , \qquad  g = \frac{u_{xy}u_{yy}}{u_y}\ .   $$
  A further solution of the constraints is
  $$  a = -u_x \ ,\qquad  f= \frac12  u_{xx}^2 \ , \qquad  g = u_{xx}u_{xy} \ .$$
  The first equation of the corresponding pair is the Hunter-Saxton equation \cite{HS}
  $$   u_{xxy } - u_x u_{xxx} =  \frac12 u_{xx}^2  $$
  and the other equation is  
  $$ u_{xyy}   -  u_x^2 u_{xxx} =   u_{xy}u_{xx} + \gamma $$
  where $\gamma$ is an arbitrary  constant.   
\end{itemize}

\section{Adler and Shabat's First Integrable Example}

In this section we look at equation (\ref{ex}), the example that Adler and Shabat gave in \cite{AS}, that is related to the potential KdV  equation.
We show that both infinite hierarchies of symmetries of this equation arise from different expansions of the same generating symmetry, but there is a critical
difference between the cases $c\not=0$ and the case $c=0$, when the equation is part of a consistent pair. 
In (\ref{ex}), by a  rescaling of $y$,  the parameter $c$ can be taken to be $0$ or $\pm1$, but it is convenient to keep it general. As noted in \cite{AS},
equation (\ref{ex}) is a potential form of  the associated Camassa-Holm equation \cite{JaCH,Hone}, 
but multiplying by $2u_y$ and differentiating with respect to $x$ we obtain the fourth order equation 
$$ u_{xxxy} - 4 u_x u_{xy} - 2 u_{xx} u_y = 0 \ .  $$
This equation is the $\partial_t=0$ reduction of the 3-dimensional  Calogero-Bogoyavlenskii-Schiff (CBS) equation \cite{CBS1,CBS2,CBS3,CBS4,CBS4a,CBS4b,CBS5,CBS6,CBS8}
$$ u_{xt} = u_{xxxy} - 4 u_x u_{xy} - 2 u_{xx} u_y\ ,  $$ 
which is a 3-dimensional generalization of the potential KdV equation, which itself is obtained
from the CBS equation by the  $\partial_x=\partial_y$ reduction.

Adler and Shabat state that for general $c$, equation (\ref{ex}) has 
two infinite hierarchies of commuting symmetries.
The first is the potential KdV hierarchy, involving only $x$-derivatives of $u$.  
The second involves $y$-derivatives and mixed derivatives with a single $x$-derivative, the first two
characteristics being 
    \begin{eqnarray*}
   Q_1   &=&  u_{xy} u_{yy} - u_y u_{xyy} + u_y^3 \ ,  \\
   Q_2   &=&  u_{yyy} - \frac32 \frac{u_{yy}^2}{u_y} + \frac{3}{2cu_y} \left( u_{xy} u_{yy} - u_y u_{xyy} + u_y^3 \right)^2 \ . 
   \end{eqnarray*}
These flows are related to the derivative nonlinear Schr\"odinger equation hierarchy. 
In the case $c=0$, differentiating (\ref{ex}) with respect to $y$ we get a total $x$-derivative,    so
\begin{equation}
  u_{xyy} = \frac{u_{yy}u_{xy}}{u_y} + u_y^2 + f(y) \ . \label{exp}   
\end{equation}
Thus when $c=0$, equations (\ref{ex}) and (\ref{exp}) form a consistent  pair. 
Also when  $c=0$ equation (\ref{ex}) is invariant under reparametrizations of 
$y$. Using this freedom in (\ref{exp}) we can take $f$ to be a constant, which we will assume from here on
(the constant can be taken to be $0$ or $\pm1$ but we will keep it general).
The pair (\ref{ex})-(\ref{exp})) is irreducible, but if we write 
$u_y = e^q$ in (\ref{exp}) it becomes  $q_{xy} = e^q + fe^{-q}$, the sine-Gordon or sinh-Gordon or Liouville equation,
depending on the choice of $f$.
The two hierarchies of symmetries of (\ref{ex}) are also symmetries of the pair 
(\ref{ex})-(\ref{exp});  the second hierarchy becomes the Schwarzian KdV hierarchy involving
only $y$-derivatives of $u$. For example, the first two 
characteristics given above, after multiplication by a suitable constant and setting $c=0$, 
become   $Q_1 = u_y$ and $Q_2 = u_{yyy} - \frac32 \frac{u_{yy}^2}{u_y}$. 
Thus far all the results presented are the results of \cite{AS}. 

The generating symmetries for the Adler-Shabat equation (\ref{ex}) are nonlocal in $u$, and can be expressed in terms of
the solutions of a number of related systems. We will use the $z$-system, to be defined below, but we could also
work in terms of B\"acklund variables or solutions of the Lax pair. The B\"acklund transformation for the Adler-Shabat equation
is that if $u$ is a solution then so is $ u - 2 v$ where $v$ satisfies 
$$ v_x = -v^2 + u_x  - \frac12 \lambda\ , \qquad   v_y = \frac1{\lambda} \left(  \frac12u_{xxy} -{u_y v_x} - {u_{xy} v}  \right) \ .   $$
Here $\lambda$ is a parameter. 
Note that we use the words ``B\"acklund variables'' here to refer to the auxiliary functions that appear in a B\"acklund
transformation. This is distinct from Adler and Shabat's usage of the term (see section 2). However a representation of the
consistent pair (\ref{ex})-(\ref{exp}) in terms of B\"acklund variables in the sense of Adler-Shabat can be found by settiing
$V = \frac{u_{xy}}{2u_y}$, in which case
$$ V_x  = -V^2 + u_x  \ , \qquad   V_y = \frac12 \left( u_y + \frac{f}{u_y} \right)\ .   $$
The first equation of this system evidently has some commonality with the first equation for the B\"acklund variable $v$ (in our
sense). 

The B\"acklund transformation for the Adler-Shabat equation (\ref{ex}) has the same superposition principle as the KdV equation.
Specifically, if the solution $u^{(1)}$ is obtained from the solution $u$ by B\"acklund transformation with the parameter $\lambda_1$, 
and the solution $u^{(2)}$ is obtained from the solution $u$ by B\"acklund transformation with the parameter $\lambda_2$, Then the solution
obtained by applying the two B\"acklund transformations to $u$, in either order, is
$$  u^{(12)} = u+\frac{\lambda_1-\lambda_2}{u^{(1)} - u^{(2)}} \ .  $$

The Lax pair for (\ref{ex}) can be obtained from the system for the B\"acklund variable by setting $v=\frac{\phi_x}{\phi} $. This gives  
\begin{equation}
    \phi_{xx} =  \left( u_x - \frac{\lambda}{2}  \right) \phi\ , \qquad 
    \phi_y   =   \frac1{2\lambda}\left( u_{xy}\phi -  2u_y\phi_x   \right)  \ .   
\end{equation}
This is actually  the Lax pair for (\ref{ex}) with $c$ an arbitrary function of $y$, but we restrict $c$ to be a constant   
as otherwise formulas become extremely lengthy. To write the generating symmetries in terms of solutions of the
Lax pair requires the use of two linearly independent solutions of the Lax pair, and the formulas can be simplified by use of a single function $z$
which is the ratio of two solutions. This satisfies what we call the $z$-system:  
  \begin{eqnarray}
    \frac{z_{xxx}}{z_x}  - \frac32 \frac{z_{xx}^2}{z_x^2} &=&  \lambda - 2 u_x \ ,  \label{zsys1}  \\
    \frac{z_y}{z_x}   = - \frac{u_y}{\lambda}\ . \label{zsys2} 
  \end{eqnarray}
Note that the $z$ system is invariant under M\"obius transformations of $z$.   

We are now ready to present the generating symmetries of (\ref{ex}). However, we first mention that  for general $c$
the equation has  4 classical  symmetries, with characteristic 
$$ \eta  =   c_1  (-2yu_y + xu_x + u)  + c_2 u_y + c_3 u_x  +  c_4  $$
where $c_1,c_2,c_3,c_4$ are constants. These correspond to 
translations of $x$, $y$ and $u$ and a scaling symmetry.   In the case $c=0$, as previously mentioned, 
the equation is also invariant under reparametrizations of $y$.  
The equation also  has 4  generating symmetries with the characteristics 
\begin{equation}
     Q = \frac{z}{z_x}, \qquad  S = \frac{z^2}{z_x}, \qquad    T  = \frac{1}{z_x} \ ,  \label{QST}
\end{equation}     
\begin{equation}
     R  =  x - \frac{2\lambda z_\lambda}{z_x}  + \frac{2yu_y}{\lambda}  \ .  \label{Req}
\end{equation}     
In the case $c=0$, then $Q,S,T$ are also generating symmetries of (\ref{exp}), the second equation of the
consistent pair. However $R$ is only a symmetry of (\ref{exp}) if $f=0$ (alternatively, when $f\not=0$,
it is necessary to define a non-trivial action of $R$ on $f$ for $R$ to be a symmetry of the full pair). 

We clain the following:  {\em Both infinite hierarchies of symmetries of equation (\ref{ex}) in the case $c\not=0$, and the
 pair (\ref{ex})-(\ref{exp}) in the case $c=0$, are obatined by expansions of $Q$ in powers of $\lambda$.} To show this
we consider the large  $|\lambda|$ expansion of $Q$, and then the small $|\lambda|$ expansion, 
first in the case $c\not=0$, and then in the case $c=0$. 

\begin{enumerate}
\item {Large $|\lambda|$ expansion}.
Writing the Schwarzian derivative of $z$ in terms of $Q=\frac{z}{z_x}$,     (\ref{zsys1}) gives
    \begin{equation}  \frac{ Q_x^2 - 2 Q Q_{xx}  -  1}{2Q^2}  =  \lambda - 2 u_x  \ .    \label{Qeq} \end{equation}
Writing   $Q = \frac{\overline{Q}}{\sqrt{-2\lambda}}$  this becomes 
    $$  \frac{ \overline{Q}_x^2 - 2 \overline{Q} \overline{Q}_{xx}}{2\overline{Q}^2} + 2 u_x =  \lambda\left( 1 -  \frac{1}{\overline{Q}^2} \right)   \  .   $$
From this we see that  $\overline{Q}$ has an expansion for large $|\lambda|$ of the form
    $$ \overline{Q} = 1   + \sum_{n=1}^\infty  \frac{\overline{Q}_n(x,y)}{ \lambda^n} \ ,  $$
with 
    \begin{eqnarray}
      \overline{ Q}_1 &=& u_x \ ,  \nonumber\\   
      \overline{ Q}_2 &=& \frac32 u_x^2 - \frac12  u_{xxx} \ , \nonumber\\    
      \overline{ Q}_3 &=& \frac52 u_x^3 - \frac54 u_{xx}^2 - \frac52 u_xu_{xxx} + \frac14 u_{xxxxx}   \ ,  \label{Qbars} \\ 
      \overline{ Q}_4 &=& \frac{35}{8} u_x^4  +  \frac{21}{8} u_{xxx}^2
       + \frac72  u_{xx}u_{xxxx} + \frac74 u_{x}u_{xxxx}
       - \frac{35}{4} u_{xxx} u_{x}^2         - \frac{35}{4} u_{xx}^2 u_{x}  
      - \frac18  u_{xxxxxxx} \ ,  
     \nonumber \\ 
      &\vdots&     \nonumber
    \end{eqnarray}   
    Each coefficient is the characteristic of  a symmetry of equation (\ref{ex}). These are the standard potential KdV flows,
    and constitute the first  infinite hierarchy of symmetries of (\ref{ex}). 
\item {Small $|\lambda|$ expansion, $c\not=0$}.   
Using the definition of $Q$ and (\ref{zsys2}) it is straightforward to check that
\begin{equation}
Q_x = \frac{ Q u_{yx} - \lambda Q_y}{u_y} \ .  \label{Qx} 
\end{equation}
This allows us to express all $x$-derivatives of $Q$ in terms of $y$ derivatives. Using this to eliminate
$x$-derivatives of $Q$ from (\ref{Qeq}) gives 
       \begin{equation}  
\left( \frac{Q_y^2}{2Q^2}  -  \frac{Q_{yy}}{Q}  + \frac{u_{yy}}{u_y} \frac{Q_y}{Q}  \right)  \frac{ \lambda^2}{u_y^2}   
+ \left( u_{xyy} - \frac{u_{xy}u_{yy}}{u_y} -u_y^2  \right) \frac{\lambda}{u_y^2}  + \frac{c}{2u_y^2}  - \frac{1}{2Q^2} = 0\ .  \label{Qid}  
       \end{equation}
Assuming  $c\not=0$  we write
       \begin{equation}  Q =  \frac{u_y}{\sqrt{c}}  \tilde{Q}       \label{Qsubs1}   \end{equation}
and find the following equation for $\tilde{Q}$:
$$  \left( \frac{\tilde{Q}_y^2}{\tilde{Q}^2}  -  \frac{2\tilde{Q}_{yy}}{\tilde{Q}}  - 2 S   \right)  \frac{ \lambda^2}{c}   
+ \frac{2\lambda M}{c} + 1 - \frac{1}{\tilde{Q}^2}  = 0    $$  
where
$$  S = \frac{u_{yyy}}{u_y} - \frac32 \frac{u_{yy}^2}{u_y^2}  \ , \qquad
    M =    u_{xyy} - \frac{u_{xy}u_{yy}}{u_y} -u_y^2 \ .     $$ 
 Evidently for small $|\lambda|$ we can find a solution of this for $\tilde{Q}$ in the form of a power series in $\lambda$
       \begin{equation}
    \tilde{Q} = 1   + \sum_{n=1}^\infty  \tilde{Q}_n(x,y) \lambda^n \ ,  \label{Qtildeexp}  
       \end{equation}
    finding the coefficients $\tilde{Q}_n(x,y)$ recursively.  
    This gives a power series solution for $Q$: 
       \begin{equation}
         Q   = \frac{u_y}{\sqrt{c}}    + \sum_{n=1}^\infty  Q_n(x,y) \lambda^n \ ,  \label{Qexp2}  
       \end{equation} 
    with
    \begin{eqnarray*}
      Q_1 &=& -\frac{Mu_y}{\sqrt{c}}\ ,  \\  
      Q_2 &=& \frac{u_y}{\sqrt{c}}\left( \frac{S}{c} + \frac32 \frac{M^2}{c^2}  \right) \ ,   \\   
      Q_3 &=& -\frac{u_y}{\sqrt{c}}\left(  \frac{M_{yy} + 3MS}{c^2}  + \frac52 \frac{M^3}{c^3}   \right) \ ,  \\
      Q_4 &=& \frac{u_y}{\sqrt{c}}\left(  \frac{3S^2 + 2S_{yy}}{2c^2}
            + \frac52  \frac{M_y^2 + 2MM_{yy} + 3M^2 S}{c^3}
            + \frac{35}{8}  \frac{M^4}{c^4}  
      \right)  \ ,    \\
      &\vdots&   
    \end{eqnarray*}
    Each coefficient is the characteristic of a symmetry of equation (\ref{ex}).  These constitue the second infinite
    hierarchy of symmetries of (\ref{ex}) in the case $c\not=0$.  
\item {Small $|\lambda|$ expansion, $c=0$}.   
  In the case $c=0$, making the substitution  (\ref{Qsubs1}) in  (\ref{Qid})
  is not appropriate.  In this case, using (\ref{exp}),  equation (\ref{Qid}) reads 
 $$
\left( \frac{Q_y^2}{2Q^2}  -  \frac{Q_{yy}}{Q}  + \frac{u_{yy}}{u_y} \frac{Q_y}{Q}  \right)  \frac{ \lambda^2}{u_y^2}   
+    \frac{\lambda f}{u_y^2}   - \frac{1}{2Q^2} = 0\ .  
 $$ 
Assuming $f\not=0$, we  make the substitution 
$$ Q =  \frac{u_y}{\sqrt{2\lambda  f}}  \tilde{Q}      $$    
to get  
$$  \left( \frac{\tilde{Q}_y^2}{2\tilde{Q}^2}  -  \frac{\tilde{Q}_{yy}}{\tilde{Q}}  - S   \right)  \frac{ \lambda}{f}   
+  1 - \frac{1}{\tilde{Q}^2}  = 0  \ .  $$  
This has a solution as an expansion in  the form (\ref{Qtildeexp}), giving the following  expansion for $Q$: 
       \begin{equation}
         Q   = \frac{u_y}{\sqrt{2 \lambda f}} \left(  1 + \frac{S \lambda}{2}
         + \frac{(3S^2 + 2S_{yy})\lambda^2}{8}
         + \frac{(5S^3 + 10S S_{yy} + 5 S_y^2 + 2 S_{yyyy} )\lambda^3}{16} + \ldots   \right)  \ .  \label{Qexpc0}  
       \end{equation} 
       We recognize that the components are the flows of the Schwarzian KdV (or UrKdV \cite{Wilson})  hierarchy.
       Furthermore, since there is no $f$ dependence in any of the components, these flows are symmetries
       also in the case $f=0$. 
\end{enumerate}  
    
Comparison of the expansion (\ref{Qexp2}) in the case $c\not=0$ and the expansion (\ref{Qexpc0}) in the case $c=0$
shows two unexpected features. First, while for $c\not=0$ it seems that $Q$ is analytic at $\lambda=0$, for $c=0$ it is
not. But also, comparing the coefficients of the expansions, it seems that in some sense there are only half as many
symmetries in the case $c=0$ as there are in the case $c\not=0$. We do not have any understanding of this.

The above discussion only shows how to obtain the two hierarchies of symmetries from the single generating symmetry $Q$.
To show they all commute it is necessary to show that the different $Q$'s associated with different values of the
parameter $\lambda$ commute. We refer the reader to \cite{RSlast} for a full discussion of how this is done for the KdV
equation, and the necessary calculation for equation (\ref{ex}) is identical. In particular, one possible choice for the
action of the symmetry $Q(\mu)$ ($Q$ with parameter $\mu$) on $z(\lambda)$ ($z$ with parameter $\lambda$) is given by
$$  \delta_{Q(\mu)}  z(\lambda) =  \frac{ z(\mu) z(\lambda)_x } { (\mu-\lambda) z(\mu)_x }\ .  $$ 

We conclude our discussion of the equation (\ref{ex}) with a discussion of the recursion relation between the components of
the two hierarchies of symmetries. Using the definition of $Q$, from equation (\ref{QST}) and equation (\ref{zsys1}), it is
straightforward to show that $Q$ satisfies the linear differential equation
$$  Q_{xxx} - 4 u_x Q_x - 2 u_{xx} Q = - 2 \lambda Q_x \ .    $$
Substituting the expansion of $Q$ for large $|\lambda|$  into this, we derive the recursion 
$$  - 2 \partial_x  \overline{Q}_{n+1}  =  (\partial_x^3 - 4 u_x \partial_x - 2 u_{xx} ) \overline{Q}_n\ ,      $$
which holds for $n\ge 0$ if we take $\overline{Q}_0 = 1$.  
Similarly, substituting the small $|\lambda|$ expansion into the differential equation for $Q$ we obtain the recursion 
$$  (\partial_x^3 - 4 u_x \partial_x - 2 u_{xx} ) {Q}_{n+1} =   - 2 \partial_x Q_{n}\ ,  $$ 
which holds for $n\ge 0$ if we take $Q_0 = u_y$.  In principle, these recursion relations can be used to find the
two hierarchies of symmetries, but it is highly nontrival to prove they are local, and the generating symmetry approach
is preferable. We note that with these defintions of $Q_0$ and $\overline{Q}_0$, the two hierachies include translations
of $x$, $y$ and $u$, but do not include the scaling symmetry, which belongs to the hierarchy associated with 
the generating symmetry $R$ that we will not discuss here.

\section{A More Intricate Example}

This section is devoted to the symmetry analysis of the new equation (\ref{ne1}), which has much in common with the analysis in
the previous section, but  has certain new features, for example the need to consider expansions of the generating symmetries 
around three values of the parameter. In (\ref{ne1}), $c$ can be an arbitrary function of $y$, but by reparametrization of $y$ it is possible
to choose $c$ to be  constant, and we shall assume this from here on. Once again, it is possible to take $c$ to have one of the values
$0,\pm1$, but we will keep $c$ to be general. 

In the case $c=0$, equation (\ref{ne1}) implies 
$$
\left(u_{xyy} - u_{y}^{2}u_{xxx} -   \frac{u_{yy}u_{xy}}{u_y}  \right)_x  = 0   \ ,   
$$ 
and thus
\begin{equation}
u_{xyy}   =  u_{y}^{2}u_{xxx} +   \frac{u_{yy}u_{xy}}{u_y}  + f(y)    \label{ne2} 
\end{equation}
for some function $f(y)$. Thus in the case $c=0$, equations (\ref{ne1})-(\ref{ne2}) form a consistent pair.
By reparametrization of $y$ it is possible to choose $f$ to be constant, and we do this from here on.
The pair is irreducible in the narrow sense described in Section 2, but
eliminating $u_{xxx}$ between equations (\ref{ne1}) and (\ref{ne2}) and substituting $u_y=e^w$, we see that in the case $c=0$ the equation
can be reduced to the second order equation  (\ref{2hyp}). 

Equation (\ref{ne1}) (and equation (\ref{ne2}) in the case $c=0$) has  classical symmetries with the characteristic 
\begin{equation}
\eta =  c_1u_y  + c_2 + c_3 x  + c_4 x^2
               + c_5 u_x + c_6 (xu_x - u) + c_7 (x^2 u_x - 2 x u) \ , 
               \label{symms}  
\end{equation}
where $c_1,\ldots,c_7$ are constants. 
In addition, in the case $c=0$, equation (\ref{ne1}) is  invariant under rescaling of $y$, and so is equation
(\ref{ne2}) if $f=0$. In greater generality (\ref{ne1}) is invariant under a joint rescaling of $y$ and $c$,
and if $c=0$, (\ref{ne2}) is invariant under a joint rescaling of $y$ and $f$. 

The Lax pair for (\ref{ne1}), with $c$ a general function of $y$,   is
\begin{eqnarray}
\psi_{xx}&=&\frac{u_{xxx}}{2(\lambda-1)}\psi,\label{LPmkdv12}\\
\psi_y&=&-\frac{u_y}{\lambda}\psi_x+\frac{u_{xy}}{2\lambda}\psi.\label{LPmkdv22}
\end{eqnarray}
The equivalent $z$-system is  
\begin{eqnarray}
 -\frac{u_{xxx}}{\lambda-1}   &=& \frac{z_{xxx}}{z_{x}} - \frac32 \frac{z_{xx}^2}{z_{x}^2}  \ ,\label{nez1} \\
   -\frac{u_{y}}{\lambda} &=&  \frac{z_{y}}{z_{x}} \ .\label{nez2}
\end{eqnarray}
Note that the $z$-system has M\"obius invariance.
The general solution of the $z$-system can be generated from any particular one by M\"obius transformations.

Equation (\ref{ne1}) (and equation (\ref{ne2}) in the case $c=0$) has  generating symmetries   
\begin{equation} 
  Q(\lambda)=\frac{z}{z_x}   , \qquad
  S(\lambda)=\frac{z^2}{z_x} , \qquad 
  T(\lambda)=\frac{1}{z_x} .    \label{Qneq}
\end{equation}
Note that $Q,S,T$ are not invariant under Mobius transformations of $z$. In fact, they transform
into each other. So in writing the above, we have assumed that we fix $z$, and are writing 
different symmetries associated with this $z$. Alternatively, it is possible to consider just $Q$
for different choices of $z$ (and later on we will do this). 
In the case $c=0$, equation (\ref{ne1}) has an additional generating symmetry 
\begin{equation} 
 R(\lambda) = \frac{z_{\lambda}}{z_x}+\frac{u}{\lambda-\lambda^2}\ .   \label{Rneq} 
\end{equation}   
If, in addition, $f=0$, then this is also a symmetry of (\ref{ne2}). In greater generality, $R$ is a symmetry
of (\ref{ne1}) for arbitrary $c$, and, if $c=0$,  of (\ref{ne2}) for arbitary $f$, if we allow changes of $c$ and $f$
given by
\begin{equation} \delta_{R(\lambda)} c  = \frac{-2c}{\lambda^2(\lambda-1)}  \  ,   \label{cact} \end{equation}   
\begin{equation} \delta_{R(\lambda)} f  = \frac{-f}{\lambda^2(\lambda-1)}    \  .   \label{fact} \end{equation}

We consider three different expansions of the generating symmetries: For small $\lambda-1$, for small
$\lambda$ and for large $\lambda$.  

\paragraph{Expansions for small $\lambda -1$.}  To obtain an expansion of $Q(\lambda)$ in powers
of $\lambda-1$ we rewrite equation (\ref{nez1}) in terms of  $Q$ to get
\begin{equation}   
\frac{u_{xxx}}{\lambda-1}  =  \frac{2QQ_{xx} - Q_x^2 +1 }{2Q^2}\ .  \label{QS} 
\end{equation}   
We can find a solution of this in the form
\begin{equation}   
  Q = \sqrt{\frac{\lambda-1}{2}}  \sum_{i=0}^\infty q_i (\lambda-1)^i
  \label{Qexp1}  
\end{equation}   
with  
\begin{eqnarray*}
q_0&=& \frac{1}{\sqrt{u_{xxx}}}  \ , \\ 
q_1&=&\frac1{4{\sqrt{u_{xxx}}}} \left(\frac{5u_{xxxx}^2}{4u_{xxx}^{3}}-\frac{u_{xxxxx}}{u_{xxx}^{2}}\right)\ , \\
q_2&=&\frac{21}{8{\sqrt{u_{xxx}}}}\left(\frac{55u_{xxxx}^4}{64u_{xxx}^{6}}-\frac{11u_{xxxx}^2u_{xxxxx}}{8u_{xxx}^{5}}
      +\frac{3u_{xxxxx}^2+4u_{xxxx}u_{xxxxxx}}{12u_{xxx}^{4}}-\frac{u_{xxxxxxx}}{21u_{xxx}^{3}}\right)\ , \\
&\vdots&   
\end{eqnarray*}
Each component in this series  is a  symmetry of (\ref{ne1}). This is the first hierarchy of local symmetries of (\ref{ne1}),
and coincides with the hierarchy of symmetries of the Hunter-Saxton equation \cite{HS} given in \cite{RS5}.  
To obtain a corresponding expansion of $R(\lambda)$ we observe that since $\frac1{Q} = (\log z)_x$, we have
$$  \frac{z_\lambda}{z} = (\log z)_\lambda  =  \int  \left(\frac1{Q}\right)_\lambda \ dx   $$
and thus
\begin{eqnarray}
  R &=&  Q \int  \left(\frac1{Q}\right)_\lambda \ dx  -  \frac{u}{\lambda(\lambda-1)}   \nonumber     \\ 
&=&  Q \int  \left(\frac1{Q}\right)_\lambda \ dx  -  \frac{u}{\lambda-1} \sum_{j=0}^\infty (-1)^j(\lambda-1)^j \nonumber \\
&=&  \sum_{i=-1}^{\infty} r_i (\lambda-1)^i   \label{Rexp1} 
\end{eqnarray}
where
\begin{eqnarray*}
r_{-1} &=& -\frac{\int\sqrt{u_{xxx}}dx}{2\sqrt{u_{xxx}}}-u\ , \\
r_{0}  &=& \frac{\int \frac{u_{xxxxx}}{u_{xxx}^{3/2}}d x}{48 \sqrt{u_{xxx}}}+\frac{5u_{xxxx}}{48 u_{xxx}^{2}}
-\frac{5 u_{xxxx}^{2}\int \sqrt{u_{xxx}}d x}{32 \left(u_{xxx}\right)^{\frac{7}{2}}}+\frac{u_{xxxxx}\int \sqrt{u_{xxx}}d x}{8 \left(u_{xxx}\right)^{\frac{5}{2}}}+u \ , \\
&\vdots&  
\end{eqnarray*}
These symmetries are apparently nonlocal, but involve a single integral. (We note that the choice of arbitrary constants in these
integrals corresponds to the freedom to add linear combinations of the $q_i$ symmetries to the $r_i$ symmetries.)  
The actions of the $r_i$ symmetries on the constant $c$ and the constant $f$ (in the case $c=0$) can be found from equations (\ref{cact}) and
(\ref{fact}): 
\begin{eqnarray*}
  \delta_{r_i} c &=& (-1)^i 2(i+2) c\ , \\  
  \delta_{r_i} f &=& (-1)^i (i+2) f \ .   
\end{eqnarray*}

\paragraph{Expansions for small $\lambda$.}  To obtain an expansion of $Q(\lambda)$ in powers of $\lambda$, we start
with the observation that if $Q$ is given as in (\ref{Qneq}) and $z$ satisfies (\ref{nez2})  then
$$  Q_x = \frac{Qu_{xy} - \lambda Q_{y}}{u_y}\ .    $$  
Using this to eliminate $x-$derivatives of $Q$ from (\ref{QS}) we obtain
\begin{eqnarray}
  0 &=&  \frac{c}{2u_y^2} - \frac{1}{2Q^2} + \frac{\lambda}{u_y^2} \left( M - \frac{c}{2}   + \frac{u_y^2}{2Q^2}  \right) 
   + \frac{\lambda^2}{u_y^2}  \left(  - \frac{Q_{yy}}{Q} + \frac{Q_y^2}{2Q^2} + \frac{Q_y}{Q}  \frac{u_{yy}}{u_y} - M - u_{xxx}u_y^2 \right)  \nonumber  \\
  &&   + \frac{\lambda^3}{u_y^2}  \left(  \frac{Q_{yy}}{Q} - \frac{Q_y^2}{2Q^2} - \frac{Q_y}{Q}  \frac{u_{yy}}{u_y} \right) \label{bigqeq} 
\end{eqnarray}
where
$$  M =  u_{xyy} - u_y^2 u_{xxx} - \frac{u_{yy}u_{xy}}{u_y}\ .   $$
For $c\not=0$, this evidently has a power series solution
\begin{equation} 
  Q = \sum_{i=0}^\infty  \overline{q}_i \lambda^i \label{powser1} 
\end{equation}  
with
\begin{eqnarray*}
  \overline{q}_0 &=& \frac{u_y}{\sqrt{c}}\ , \\
  \overline{q}_1 &=& -\frac{Mu_y}{c^{3/2}}\ ,     \\
  \overline{q}_2 &=& \frac{u_y}{c^{3/2}}\left(S + u_y^2 u_{xxx} + \frac{3M^2}{2c}  \right)  \ ,   \\
  \overline{q}_3 &=& \frac{u_y}{c^{3/2}}\left(u_y^2 u_{xxx} - \frac{3u_y^2u_{xxx}M + 3 M S + M_{yy}}{c}   -  \frac{5M^3}{2c^2}  \right)  \ ,     \\
  &\vdots&   
\end{eqnarray*}
where
$$ S =  \frac{u_{yyy}}{u_y} - \frac32 \frac{u_{yy}^2}{u_y^2}\ .     $$
For $c=0$, we note (looking at (\ref{ne2})) that the quantity $M$ introduced above is equal to the constant $f$,
and we write $f=k^2$. Assuming $k$ is nonzero, the power series solution of (\ref{bigqeq}) now takes the form 
\begin{equation}
  Q =  \frac1{\sqrt{2\lambda}}\sum_{i=0}^\infty  \overline{q}_i \lambda^i\ ,  \label{powser2} 
\end{equation}    
where the first few coefficients are given by  
\begin{eqnarray*}
  \overline{q}_0 &=& \frac{u_y}{k}\ , \\
  \overline{q}_1 &=& \frac{u_y}{2k^3} \left( S + u_y^2 u_{xxx} \right)    ,     \\
  \overline{q}_2 &=& \frac{u_y}{8k^5} \left(  10 u_{xy}^2 u_{xxx} u_y^2  + 10 u_{xy} \left( u_y^3u_{xxxx} - 2 u_{xxx}u_y u_{yy}  \right)  
  + u_y^4 \left(2 u_{xxxxx}  - 5 u_{xxx}^2   \right)   \right. \\
  && \left.   + 10 u_{yy}^2 u_{xxx} - 10 u_{xxxx} u_{yy} u_y^2    + 10 u_y^2 u_{xxx} S + 3 S^2 + 2 S_{yy}   \right)   \ ,    \\
  &\vdots&   
\end{eqnarray*}
The coefficients given explicitly here give the false impression that $\overline{q}_i$ is homogeneous as a function of $k$.
But this is not the case. The coefficent $\overline{q}_3$, which is far too lengthy to be given explicitly here,  has
terms proportional to both  $k^{-5}$ and $k^{-7}$.  However it is true that most singular terms in $\overline{q}_i$ are proportional
to $k^{-2i-1}$ and this gives a clue how to take the $k\rightarrow 0$ limit. Instead of looking at $Q(x,y,\lambda)$ we look at $\tilde{Q}(x,y,\mu)$
given by
$$   Q(x,y,\lambda) = \frac{1}{\sqrt{2 \lambda} k} \tilde{Q}(x,y,\mu) $$
where $\mu = \frac{\lambda}{k^2} $.   $\tilde{Q}$ satisfies the equation 
$$
   \frac{1}{u_y^2}  - \frac{1}{\tilde{Q}^2}   + \frac{\mu}{u_y^2}  
   \left(  - \frac{\tilde{Q}_{yy}}{\tilde{Q}} + \frac{\tilde{Q}_y^2}{2\tilde{Q}^2} + \frac{\tilde{Q}_y}{\tilde{Q}}  \frac{u_{yy}}{u_y}  -  \frac{\mu u_{xxx}u_y^2}{1-\mu k^2} \right)  
  = 0 \ ,
$$
  and has a power series expansion
\begin{equation} \tilde{Q} =  \sum_{i=0}^\infty  \tilde{q}_i \mu^i\ ,  \label{powser3} \end{equation}
  with $\tilde{q}_i = k^{2i+1} \overline{q}_i$.  The symmetries  $\overline{q}_i$ do not have a limit as $k\rightarrow 0$, but the
  symmetries $\tilde{q}_i$ do.   The coefficients $\tilde{q}_0, \tilde{q}_1, \tilde{q}_2$, which are valid in the $k\rightarrow 0$ limit,  can be
  obtained from the formulas given above for $\overline{q}_0,\overline{q}_1,\overline{q}_2$ 
  in the  $k\not=0$ case, simply by setting $k=1$.  
  
  To summarize: although the expansions are different for the cases  $c\not=0$ (equation \ref{powser1}),
  $c=0$ and $f\not=0$ (equation \ref{powser2}), and $c=f=0$ (equation \ref{powser3}), in each
  case there is an expansion of $Q$ for small $\lambda$, giving a second hierarchy of local symmetries of (\ref{ne1}). 
In each case the lowest order symmetry in the hierarchy is  simply $u_y$ (which, we recall, is one of the point symmetries of (\ref{ne1}),
see (\ref{symms})).  

To find the corresponding expansions of $R$  we use (\ref{Rneq}), for which we first need to find an expression for $z$.
We know $Q = \frac{z}{z_x}$, from which
\begin{equation}  \log z =  \int \frac{dx}{Q}  + C(y,\lambda) \ ,  \label{zeq}  \end{equation} 
where $C(y,\lambda)$ is a currently undetermined function of $y$ and $\lambda$. To satisfy  (\ref{nez2}) we  need
$$ \frac{u_y}{\lambda} =  Q  \left( \int \frac{Q_y}{Q^2} \ dx -    C_y \right)   \ .     $$ 
In the case $c\not=0$,  from  (\ref{powser1}),  we have $Q = \frac{u_y}{\sqrt{c}} + O(\lambda)$ for small $\lambda$,
and the first term on the right hand side is regular at $\lambda=0$. It follows that $C_y$ cannot be regular at $\lambda=0$, 
and must have a singular part $-\frac{\sqrt{c}}{\lambda}$.  In fact, without loss of generality, we can take 
$$    C = -\frac{\sqrt{c}y}{\lambda}  $$
as the regular part of $C_y$ at $\lambda=0$ can be absorbed into the choice of limits in the integral, and adding a function
of $\lambda$ alone to $C$ corresponds to adding some ($\lambda$-dependent) multiple of $Q$ to $R$. Thus, using 
(\ref{Rneq}), we have
\begin{eqnarray}
  R &=& \frac{z_\lambda/z}{z_x/z} +\frac{u}{\lambda-\lambda^2}  \nonumber\\
  &=&   Q \left(  \int \left(\frac1{Q}\right)_\lambda\ dx  + \frac{\sqrt{c}y}{\lambda^2} \right)  + u \sum_{i=-1}^\infty \lambda^i  \nonumber\\
  &=&  \sum_{i=-2}^\infty  \overline{r}_i \lambda^i     \label{Rexp2}    
\end{eqnarray}
where
\begin{eqnarray*}
  \overline{r}_{-2} &=& yu_y \ , \\
  \overline{r}_{-1} &=& u - \frac{yMu_y}{c} \ ,  \\
  \overline{r}_{0} &=&  u + \sqrt{c} y \overline{q}_2 + \frac{u_y}{c^2} \int Mu_y \ dx \ ,   \\
  &\vdots&
\end{eqnarray*}
Apart from the first two, these symmetries are nonlocal, involving a single integral.

Moving to the analysis of $R$ in the case $c=0$, the appropriate form of $C$ to
take in equation (\ref{zeq}) in order to satisfy (\ref{nez1}) is 
$$  C = -\frac{\sqrt{2}ky}{\sqrt{\lambda}}\ .   $$
Using this  we obtain
$$ R = \sum_{i=-2}^\infty  \overline{r}_i \lambda^i     $$
with 
\begin{eqnarray*}
  \overline{r}_{-2} &=& \frac12 yu_y \ , \\
  \overline{r}_{-1} &=& u +  \frac12 k y \overline{q}_1  + \frac12 u_y \int \frac{dx}{u_y}  \ ,  \\
  \overline{r}_{0} &=& u +   \frac12 k y \overline{q}_2 + \frac12 \overline{q}_1 \int \frac{dx}{u_y}  - \frac32 k u_y  \int \frac {\overline{q}_1}{u_y^2}\ dx     ,  \\
  &\vdots&
\end{eqnarray*}

\paragraph{Expansions for large $\lambda$.}   It is in fact more convenient to consider this expansion
as an expansion in powers of $\lambda-1$.  We start with the $z$-system (\ref{nez1})-(\ref{nez2}). For large
$\lambda$  the left-hand sides are both zero, so to leading order we have
$$  z = \frac{\alpha x + \beta}{\gamma x + \delta}\ . $$
where $\alpha,\beta,\gamma,\delta$ are constants, with $\alpha\delta-\beta\gamma=1$ without loss of generality. 
(The solution of (\ref{nez1})-(\ref{nez2}) with vanishing left hand side takes this form with $\alpha,\beta,\gamma,\delta$
all functions of $\lambda$. We assume these functions tend to constants in the large $\lambda$ limit, and these are
the constants indicated above. Further constants of integration appear at each stage of constructing
the large $\lambda$ expansion; these can all be handled by treating $\alpha,\beta,\gamma,\delta$ as
polynomials in $\lambda-1$, to the relevant order.)  It is then convenient to take the expansion of $z$ in the form
$$
z =  \frac{\alpha x + \beta}{\gamma x + \delta}    + \frac{1}{(\gamma x+ \delta)^2}  \sum_{i=1}^\infty  \frac{ z_i(x,y) }{ (\lambda-1)^i}  \ . 
$$
Substituting in (\ref{nez1})-(\ref{nez2}) we obtain
\begin{eqnarray*}
  z_1 &=& -u\ ,  \\
  z_2 &=&  \tilde{z}_2   - \frac{\gamma u^2}{\gamma x + \delta} \ ,  \\
  z_3 &=&  \tilde{z}_3   + \frac{2 \gamma u \tilde{z}_2}{\gamma x + \delta}   - \frac{\gamma^2 u^3}{(\gamma x + \delta)^2} \ ,  \\
  &\vdots& 
\end{eqnarray*}   
where
\begin{eqnarray*}
\left(\tilde{z}_2\right)_{xxx}  &=&  u_x u_{xxx} + {\textstyle{\frac32}} u_{xx}^2  \ ,   \\  
\left(\tilde{z}_3\right)_{xxx}  &=&  -3 u_{xx} \left(\tilde{z}_2\right)_{xx} - u_{xxx} \left(\tilde{z}_2\right)_{x} + {\textstyle{\frac32}} u_xu_{xx}^2 \ , \\   
  &\vdots&   \\   
\left(\tilde{z}_2\right)_{y}  &=&  u_y(1+u_x)     \ , \\
\left(\tilde{z}_3\right)_{y}  &=&  -u_y\left( 1 + u_x + \left( \tilde{z_2} \right)_x \right)    \ ,  \\   
  &\vdots& 
\end{eqnarray*}   
Substituting the expansion for $z$ into the formula in (\ref{Qneq}) for  $Q$ we obtain an expansion for $Q$, depending on the parameters
$\alpha,\beta,\gamma,\delta$. Linearizing in $\beta,\gamma$ we obtain
$$ Q =  Q_1 + \frac{Q_2\beta}{\alpha} + \frac{Q_3 \gamma}{\delta}  + \ldots   $$ 
where
\begin{eqnarray}
  Q_1 &=&  x + \frac{xu_x - u}{\lambda-1}  + \frac{x u_x^2 - uu_x - x (\tilde{z}_2)_x + \tilde{z}_2}{(\lambda-1)^2}  + \ldots \ ,  \label{QQser1}\\   
  Q_2 &=&  1 + \frac{u_x}{\lambda-1}  +\frac{ u_x^2 - (\tilde{z}_2)_x }{(\lambda-1)^2}     + \ldots   \ , \label{QQser2}\\
  Q_3 &=&  x^2 + \frac{x^2u_x - 2xu}{\lambda-1} + \frac{x^2 u_x^2 - 2xuu_x + u^2 - x^2 (\tilde{z}_2)_x + 2x\tilde{z}_2 }{(\lambda-1)^2}   + \ldots  \ .  \label{QQser3}
\end{eqnarray}
We see in the first few terms 6 of the 7 generalized point symmetries given in (\ref{symms}). The higher terms, involving $\tilde{z}_2, \tilde{z}_3,\ldots$
are nonlocal, though the simplest one (associated with the third term in $Q_2$),  $u_t = u_x^2 - (\tilde{z}_2)_x$,  can be differentiated twice to give 
$u_{txx}   =  u_x u_{xxx} + \frac12 u_{xx} ^2 $ which is the Hunter-Saxton equation \cite{HS}.  

What we have done in the previous paragraph is to compute $Q$ for the most general solution of the $z$-system.
As mentioned previously, this is equivalent to computing the 3 generating symmetries $Q,S,T$ for a single
solution of the $z$-symmetry. The symmetries that appear as the coefficients of different powers of $\frac{1}{\lambda-1}$ in
a single expansion commute with each other. However symmetries appearing in different expansions do not
necessarilly commute with each other (see \cite{RSlast}). 

Moving now to the corresponding expansion for $R$, we note that the formula (\ref{Rneq}) for $R$ is invariant
under Mobius transformations of $z$, and thus there is only a single expansion. This takes the form
\begin{equation}
  R
  =   \frac{uu_x+u -2\tilde{z}_2  }{(\lambda-1)^3}  
  +  \frac{uu_x^2 - u - u(\tilde{z}_2)_x - 2 u_x \tilde{z}_2 - 3 \tilde{z}_3   }{(\lambda-1)^4}   
  + \ldots \ .  \label{RRser}
\end{equation}  

\paragraph{Summary.} The following picture emerges. The generating symmetry $Q$ has distinct expansions for small $\lambda-1$, small $\lambda$
and large $\lambda$. For small $\lambda-1$ there is a single expansion (\ref{Qexp1}), valid for all values of the constants $c$ and $f$,  giving an infinite
hierarchy of local symmetries.  For small $\lambda$, there is a single expansion, but this takes different forms in the three cases $c\not=0$ (equation (\ref{powser1})),   
$c=0,f\not=0$ (equation (\ref{powser2})), and    $c=f=0$ (equation (\ref{powser3})).  In each case there is an infinite hierarchy of local symmetries, with the
lowest order symmetry corresponding to $y$-translation, one of the classical symmetries. When $c=0$ there appear, in some sense, to only be half as many
of these symmetries as when $c\not=0$. Finally, for large $\lambda$, there are three distinct expansions, (\ref{QQser1}), (\ref{QQser2}) and (\ref{QQser3}),
valid for all values of the constants $c$ and $f$.
In each of these expansions only the two lowest order symmetries are local, and these give the remaining six classical symmetries given in (\ref{symms}).  

Before summarizing the results for the generating symmetry $R$, we recall that this is only a symmetry in the standard sense of (\ref{ne1}) 
if $c=0$, and of (\ref{ne2}) if in addition $f=0$. However, if we allow $R$ to also act on the constants $c$ and $f$ then it is a symmetry for
arbitrary values of $c$ and $f$.  For small $\lambda-1$, $R$ has the expansion (\ref{Rexp1}), in which all terms are nonlocal, involving a single
integral. For small $\lambda$, $R$ has an expansion of the form (\ref{Rexp2}), in which the coefficients are different depending on whether $c\not=0$ or
$c=0$; in the latter case the only local symmetry in the hierarchy is $y$-rescaling, but in the former there is a second local symmetry. All other symmetries in the
hierarchy involve a single integral. Finally, for large $\lambda$ there is also only a single expansion of $R$, equation (\ref{RRser}), with all terms nonlocal. 

\paragraph{Recursion operators}
We mention in conclusion of our symmetry analysis for equation (\ref{ne1}) that recursion formulae between the different
components of the different hierarchies of symmetries can be obtained from the identities 
\begin{eqnarray*}
  Q_{xxx} &=&  \frac{1}{\lambda-1} \left( u_{xxxx} Q + 2 u_{xxx} Q_x   \right) \ ,   \\
  R_{xxx} &=&  \frac{1}{\lambda-1} \left( u_{xxxx} R + 2 u_{xxx} R_x  + \frac{1}{\lambda(\lambda-1)} \left( u_{xxx} + uu_{xxxx} + 2u_x u_{xxx}  \right)    \right) \ .    
\end{eqnarray*}
We give just two examples.  Substituting the small $\lambda-1$ expansion (\ref{Qexp1}) into the $Q$-identity, we obtain the recursion
$$  ( 2 u_{xxx}\partial_x + u_{xxxx})  q_{i+1}  =   q_{i,xxx}  \ , \qquad  i=0,1,\ldots  $$   
Substituting the small $\lambda$ expansion in the case $c\not=0$, equation (\ref{powser2}), into the $Q$-identity, we obtain the recursion
$$  ( \partial_x^3  + 2 u_{xxx}\partial_x + u_{xxxx})  \overline{q}_{i+1}  =   \overline{q}_{i,xxx}  \ , \qquad  i=0,1,\ldots  $$   

\section{Concluding remarks}

The research presented in this paper very much reflects the problem that although we know that a necessary property for integrability of a PDE is 
the existence of an infinite number of symmetries, and this can be used to search for integrable equations, it is much harder to know when we have
found all the symmetries of a given equation or whether a given infinite set is sufficient.

We have shown that different infinite hierarchies of symmetries for a given equation can be obtained by different expansions, around different
values of the parameter, of a single generating symmetry. This suggests to us that the existence of a generating symmetry  (or symmetries) for an
equation is fundamental to integrability --- but we still have no idea how many generating symmetries there should be, nor (since they are written
in terms of some nonlocal variables) in what form to search for them. Nevertheless, a routine effort should be made to identify generating symmetries,
and we think this may well be easier than, for example, looking for recursion operators. 

Also in this paper we have  extended Adler and Shabat's notion of a consistent pair of third order equations. Although the only application we
know so far of this kind of system is \cite{DS}, it shows how even one of the most basic notions in differential equations --- that of the
order of a differential equation --- is not as simple as it might seem. In both examples that we have studied in depth we have seen how the
symmetry structure changes when the value of the parameter $c$ is set to $0$ and the equation becomes part of a consistent pair (and an extra
scaling symmetry appears). In particular, the type of expansions needed to derive hierarchies from the generating symmetry changes, and we have no
understanding of why this happens. A more detailed study of the limit $c\rightarrow 0$ would certainly be interesting.


\begin{thebibliography}{10}
  
\bibitem{AS} 
{\sc Adler, V.~E., and Shabat, A.~B.}
\newblock Toward a theory of integrable hyperbolic equations of third order. 
\newblock {\em J. Phys. A 45}  (2012), 385207.

\bibitem{CBS2} 
{\sc Bogoyavlenskii, O. I.}
\newblock  Breaking solitons in new two-dimensional integrable equations.
\newblock {\em Math. USSR Izv. 34} (1990), 245--259. 

\bibitem{CBS3} 
{\sc Bogoyavlenskii, O. I.}
\newblock  Breaking solitons, II.  
\newblock {\em Math. USSR Izv. 35} (1990), 245--248. 

\bibitem{CBS4b} 
{\sc Bogoyavlenskii, O. I.}
\newblock  Breaking solitons in 2+1-dimensional integrable equations. 
\newblock {\em Russ. Math. Surveys 45}, number 4 (1990), 1--86.

\bibitem{CBS4} 
{\sc Bogoyavlenskii, O. I.}
\newblock  Breaking solitons, III.  
\newblock {\em Math. USSR Izv. 36} (1991), 129--137. 

\bibitem{CBS4a} 
{\sc Bogoyavlenskii, O. I.}
\newblock  Breaking solitons, IV.  
\newblock {\em Math. USSR Izv. 37} (1991), 475--487. 

\bibitem{CBS8} 
{\sc Bruz\'on, M.~S.,  Gandarias, M.~L., Muriel, C., Ram\'irez, J.,  Saez, S., and Romero, F.~R. }
\newblock  The Calogero-Bogoyavlenskii-Schiff equation in 2+1 dimensions. 
\newblock {\em Theo.  Math. Phys., 137} (2003), 1367--1377. 

\bibitem{DS}
{\sc Demskoi, D.~K., and Schief, W.~K.}
\newblock  On steady motions of an ideal fibre-reinforced fluid in a curved stratum. {G}eometry and integrability.
\newblock  {\em J. Phys. A 54} (2021), 505205.   

\bibitem{Dic}
{\sc Dickey, L.~A.}
\newblock {\em Soliton equations and {H}amiltonian systems}, second~ed.,
  vol.~26 of {\em Advanced Series in Mathematical Physics}.
\newblock World Scientific Publishing Co. Inc., River Edge, NJ, 2003.
  
\bibitem{cp9} 
{\sc Habibullin, I.~T., and Khakimova. A.~R.}
\newblock  Invariant manifolds of hyperbolic integrable equations and their applications.
\newblock  {\em J. Math. Sci 257} (2021), 410--423. 

\bibitem{HK} 
{\sc Habibullin, I.~T., and Khakimova. A.~R.}
\newblock  A Direct Algorithm for Constructing Recursion Operators and Lax Pairs for Integrable Models.
\newblock  {\em Theo.  Math. Phys. 196} (2018) 1200--1216. 
   

\bibitem{Hone}
{\sc Hone, A.N.W.}
\newblock  The associated {C}amassa-{H}olm equation and the {K}d{V} equation.  
\newblock  {\em  J. Phys. A 32} (1999), L307-314. 

\bibitem{HS}  
{\sc Hunter, J.~K., and Saxton, R.}
\newblock Dynamics of director fields,
\newblock {\em SIAM J. Appl. Math. 51} 6 (1991), 1498--1521.     

\bibitem{addsym1}
{\sc Ibragimov, N.~H., and \v{S}abat, A.~B.}
\newblock The {K}orteweg-de\thinspace {V}ries equation from the group  standpoint.
\newblock {\em Dokl. Akad. Nauk SSSR 244} (1979), 57--61.

\bibitem{addsym2}
{\sc Khor'kova, N.~G.}
\newblock Conservation laws and nonlocal symmetries.
\newblock {\em Mat. Zametki 44} (1988), 134--144.

\bibitem{cp5}  
{\sc Kuznetsova, M.~N., Pekcan, A., and Zhiber, A.~V.}
\newblock The {K}lein-{G}ordon equation and differential substitutions of the form $v =\phi(u,u_x,u_y)$.
\newblock {\em SIGMA 8} (2012), 090.

\bibitem{cp8} 
{\sc Lou, S.}
\newblock Twelve sets of symmetries of the {C}audrey-{D}odd-{G}ibbon-{S}awada-{K}otera equation .
\newblock  {\em Phys. Lett. A 175} (1993), 23--26. 

\bibitem{lou2012}
{\sc Lou, S., Hu, X. and Chen,   Y.} 
\newblock Nonlocal symmetries related to {B}{\"a}cklund transformation and their applications.
\newblock {\em J. Phys. A 45} (2012) 155209.

\bibitem{cp4} 
{\sc Meshkov, A.~G., and Sokolov, V.~V.} 
\newblock  Hyperbolic equations with third order symmetries. 
\newblock  {\em Theo. Math. Phys. 166} (2011), 43--57. 

\bibitem{cp7} 
{\sc Meshkov, A.~G., and Sokolov, V.~V.} 
\newblock  Vector hyperbolic equations on the sphere possessing integrable third-order symmetries.
\newblock  {\em Lett. Math. Phys. 104}  (2014),  341--360.  

\bibitem{cp3} 
{\sc Mikhailov, A.~V., Novikov, V.~S., and Wang, J.~P.}
\newblock  On classification of integrable non-evolutionary equations.
\newblock  {\em Stud. Appl. Math 118} (2007), 419--457. 

\bibitem{book1} 
{\sc Mikhailov, A.~V.,  Shabat, A.~B., and Sokolov, V.~V.}
\newblock  The symmetry approach to classification of integrable equations.
\newblock  In {\em What is Integrability?}, ed. Zakharov, V.~E., Springer--Verlag (1991). 

\bibitem{book2} 
{\sc Mikhailov, A.~V.,  and Sokolov, V.~V.}
\newblock Symmetries of differential equations and the problem of integrability. 
\newblock  In {\em Integrability}, ed. Mikhailov, A.~V., Springer (2009). 

\bibitem{MGK}
{\sc Miura, R.~M., Gardner, C.~S.,  and Kruskal, M.~D.}
\newblock Korteweg-de {V}ries equation and generalizations. {II}. {E}xistence of conservation laws and constants of motion .
\newblock {\em J. Math. Phys. 9} (1968), 1204--1209.   

\bibitem{cp1}
{\sc Novikov, V.}
\newblock Generalizations of the Camassa-Holm equation.
\newblock  {\em J. Phys. A 42} (2009), 342002. 

\bibitem{cp2}   
{\sc Novikov, V.~S., and Wang, J.~P.}
\newblock  Symmetry structure of integrable non-evolutionary equations. 
\newblock  {\em Stud. Appl. Math 119} (2007), 393--428. 

\bibitem{OEVEL1998161}
{\sc Oevel, W. and Carillo, S.}
\newblock Squared eigenfunction symmetries for soliton   equations: {P}art {i}.
\newblock {\em J. Math. Anal. Appl 217} (1998) 161--178.


\bibitem{Ol1}
{\sc Olver, P.~J.}
\newblock Evolution equations possessing infinitely many symmetries.
\newblock {\em J. Math. Phys. 18}  (1977), 1212--1215.

\bibitem{addsym3}
{\sc Orlov, A.~Y., and Schulman, E.~I.}
\newblock Additional symmetries for integrable equations and conformal algebra   representation.
\newblock {\em Lett. Math. Phys. 12} (1986), 171--179.
  
\bibitem{Sashalast}
{\sc Rasin, A.~G.}
\newblock Computation of generating symmetries.
\newblock {\em Comm. Nonlinear Sci. and Num. Simul. 118} (2023), 107003. 

\bibitem{RSG}
{\sc Rasin, A.~G., and Schiff, J.}
\newblock The {G}ardner method for symmetries.
\newblock {\em J. Phys. A 46} (2013), 155202.

\bibitem{RS5}
{\sc Rasin, A.~G., and Schiff, J.}
\newblock B\"{a}cklund transformations for the {C}amassa-{H}olm equation.
\newblock {\em J. Nonlinear Sci. 27} (2017), 45--69.

\bibitem{RS6}
{\sc Rasin, A.~G., and Schiff, J.}
\newblock B\"{a}cklund transformations for the {B}oussinesq equation and
  merging solitons.
\newblock {\em J. Phys. A 50} 32 (2017), 325202.

\bibitem{RS7}
{\sc Rasin, A.~G., and Schiff, J.}
\newblock Unfamiliar aspects of {B}\"{a}cklund transformations and an
  associated {D}egasperis-{P}rocesi equation.
\newblock {\em Teoret. Mat. Fiz. 196} 3 (2018), 449--464.

\bibitem{RS8} 
{\sc Rasin, A.~G., and Schiff, J.}
\newblock A simple-looking relative of the {N}ovikov, {H}irota-{S}atsuma and {S}awada-{K}otera equations. 
\newblock {\em J. Nonlin. Math. Phys.} 26 (2019), 555--568.   
  
\bibitem{RSlast}
{\sc Rasin, A.~G., and Schiff, J.}
\newblock Four symmetries of the {K}d{V} Equation. 
\newblock {\em J. Nonlin. Sci. 32} (2022), 68.    

\bibitem{CBS1} 
{\sc Schiff, J.}
\newblock  Integrability of Chern-Simons-Higgs vortex equations and a reduction of the self-dual Yang-Mills
equations to three dimensions. 
\newblock in {\em Painlev\'e Trascendents, Their Asymptotics and Physical Applications}, ed. Levi, D., et al., Plenum (1992). 

\bibitem{JaCH}
{\sc Schiff, J.}
\newblock The {C}amassa-{H}olm equation: A loop group approach.  
\newblock {\em Physica D 121} (1998), 24--43.

\bibitem{cp6} 
{\sc Startsev, S.~Ya.}
\newblock Formal integrals and {N}oether operators of nonlinear hyperbolic partial differential systems admitting a rich set of symmetries.
\newblock  {\em SIGMA 13} (2017), 034. 

\bibitem{CBS6} 
{\sc Toda, K., Yu, S.-J, and Fukuyama, T.}
\newblock   The Bogoyavlenskii-Schiff hierarchy  and integrable equations in (2+1) dimensions. 
\newblock {\em Rep. Math. Phys. 44} (1999), 247--254.
    
\bibitem{Wilson} 
{\sc Wilson, G.}
\newblock  On the quasi-hamiltonian formalism of the {K}d{V} equation.
\newblock {\em Phys. Lett. A 132} (1988), 445--450.

\bibitem{CBS5} 
{\sc Yu, S.-J,, Toda, K., Sasa, N., and Fukuyama, T.}
\newblock  $N$ soliton solutions to the Bogoyavlenskii-Schiff equation and a quest for the soliton solution in (3+1) dimensions. 
\newblock {\em J. Phys. A 31} (1998), 3337--3347. 

\end{thebibliography}
\end{document}